\documentclass[11pt,twoside]{article}
\usepackage{./asp2010}

\resetcounters

\begin{document}

\title{Massive non-thermal radio emitters: new data and their modelling}
\author{Delia~Volpi,$^1$ Ronny~Blomme,$^1$ Michael~De~Becker$^{2,3}$, and Ya\"el~Naz\'e$^2$
\affil{$^1$Royal Observatory of Belgium, Ringlaan 3, B-1180 Brussel, Belgium\\
       Delia.Volpi@oma.be\\
$^2$Institut d'Astrophysique, Universit\'{e} de Li\`ege, All\'{e}e du 6 Ao\^{u}t, 17,
B\^{a}t B5c, B-4000 Li\`ege (Sart-Tilman), Belgium\\
$^3$Observatoire de Haute-Provence, F-04870 Saint-Michel L'Observatoire, France}
}

\begin{abstract}
 During recent years some non-thermal radio emitting OB stars have been discovered to be binary, or multiple systems. The non-thermal emission is due to synchrotron radiation that is emitted by electrons accelerated up to high energies. The electron acceleration occurs at the strong shocks created by the collision of radiatively-driven winds. Here we summarize the available radio data and more recent observations for the binary Cyg OB2 No. 9. We also show a new emission model which is being developed to compare the theoretical total radio flux and the spectral index with the observed radio light curves. This comparison will be useful in order to solve fundamental questions, such as the determination of the stellar mass loss rates, which are perturbed by clumping. 
\end{abstract}

\section{Introduction}
Several massive stars, of both Wolf-Rayet and OB-type, show evidence of non-thermal radio emission \citep{bieging1989}. This non-thermal radiation is produced by relativistic electrons accelerated at collisionless shocks through the Fermi mechanism. Following the theoretical model of \citet{eichler1993} the shocks are created by the strong radiatively-driven winds that collide in the binary or multiple systems. This model was confirmed by the observations of WR140 \citep{Dougherty2005}. The accelerated electrons advect away cooling down due to adiabatic and inverse Compton losses. In the presence of the magnetic field, they emit synchrotron radiation in the radio band. The synchrotron radiation is partially absorbed by the free-free absorption mechanism, leading to flux variability locked to the orbital phase. This variability consists of two contributions: the intrinsic synchrotron variations due to the changes in collision energy linked to the orbital phase and the changing free-free absorption along the line of sight.

\begin{figure}
\centering
\includegraphics[bb= +45 0 481 226, scale=0.85]{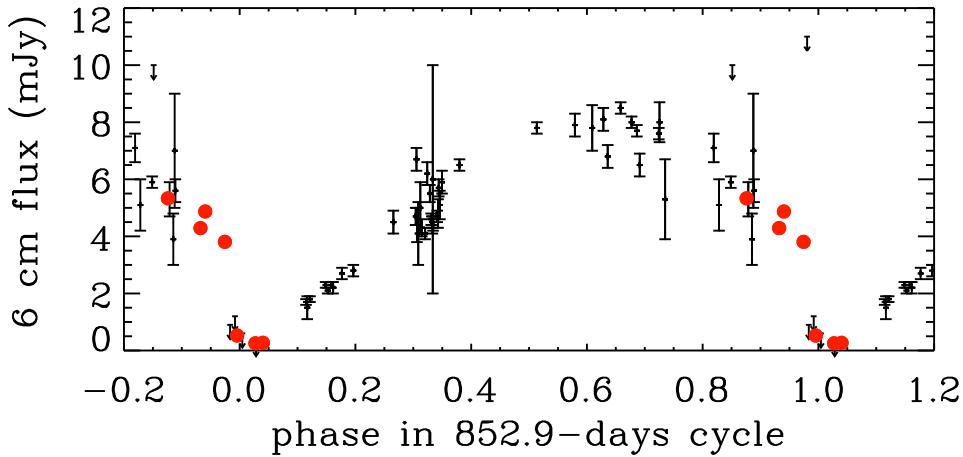}
\caption{Cyg OB2 No. 9: $6$~cm flux versus phase during the complete orbital period of $852.9$~days. The black points are the available VLA radio data from \citet{vanloo2008}, the red ones are our preliminary reductions of the recent observations from EVLA. A sharp drop during periastron passage is clearly observable.}\label{fig1}
\end{figure}

Recently new observations for the O5 + O6-7 binary Cyg OB2 No. 9 were obtained. Here these new radio data together with previously published ones are shown (Fig.~\ref{fig1}). A new emission model, which is applicable both to Wolf-Rayet and OB stars, is presented. This model is being developed in order to compare the observations with the theoretical results. The comparison is essential to investigate important issues related to massive stars astrophysics, such as the clumping effect on the mass loss rate determination of single stars, the particle acceleration mechanism at the shocks and the binary frequency of early-type stars.

\section{Observations: Cyg OB2 No. 9}
\citet{vanloo2008} discovered that the observed VLA radio fluxes of Cyg OB2 No. 9 at $3.6$, $6$, and $20$~cm show recurrent variations with a period of $\approx 2.355$~years. They accordingly suggested that the star might be a binary system. Binarity was established soon afterwards by the discovery of line splitting in the optical \citep{naze2008} and a first orbital solution of Cyg OB2 No. 9 was obtained by \citet{naze2010}, giving a period of $852.9$~days. The $6$~cm radio fluxes from \citet{vanloo2008} are shown in Fig.~\ref{fig1} as black points. The phase-locked flux variability, which is a fingerprint of non-thermal emission in a binary, is clearly observable. The measurements cover a period of about $20$~years but with only a few data points around periastron passage. 

 In the framework of a large multi-wavelength campaign initiated by one of us (YN) with the goal to monitor the 2011 periastron passage of Cyg OB2 No. 9, the first observable since the discovery of binarity, we obtained new EVLA data.
They are superimposed on the old ones as red points (the data reduction is only preliminary). The new data agree very well with the old ones and they show a very sharp drop just before periastron.

\section{Modelling Cyg OB2 No. 9}
 A previous version of the emission code was developed and applied to the binary systems Cyg OB2 No. 8A \citep{blomme2010} and No. 9 \citep{volpi2011a,volpi2011b}. The code presents several approximations: the input hydrodynamical variables were determined analytically, the orbital motion was not included, and the shocks were considered coincident with the contact discontinuity. The strength of the shock was chosen to have a constant value. These approximations created a discrepancy between our $6$~cm light curves, and the corresponding observations. In the case of Cyg OB2 No. 9 our fluxes are too high and the maximum occurs too early.

A new emission code is now being developed in order to solve the above discrepancy. The input hydrodynamical variables are obtained with the Athena code, evolving the gas hydrodynamics and including the orbital motion.

The shock positions are found in the hydrodynamical results by looking for the density jump. The shock strengths are now determined point by point leading to a more consistent model for the particle acceleration. The electrons accelerated at the shocks are followed along the post-shock streamlines while they are moving away and cooling down due to the adiabatic and inverse Compton losses. The relativistic electrons in the presence of the local magnetic field emit radio synchrotron emission. The magnetic field is locally estimated following \cite{blomme2010}. The total flux is then calculated including free-free absorption and emission. 

\section{Future work}
The phase-locked variability in the total flux will be analyzed and compared with the old VLA and new EVLA data from Cyg OB2 No. 9. The space of the model parameters will be better investigated in order to improve our physical understand of colliding-wind binaries. The code will also be applied to other massive binaries and multiple systems. The complementarity between observations in different wavebands as optical and X-rays will provide relevant constraints on the modelling of Cyg OB2 No. 9. This will help us to further understand the production of non-thermal radio emission in this system.
\acknowledgements 
\begin{sloppypar} DV is supported by contract Action 1 project MO/33/024 (Colliding winds in O-type binaries). MdB and YN acknowledge support from FNRS, CFWB, the PRODEX contracts, and the ARC. We thank Joan Vandekerckhove for his help with the reduction of the EVLA data.\end{sloppypar}

\bibliography{aspvolpi}

\end{document}